\documentclass[journal]{IEEEtran}

\usepackage{amsmath}
\usepackage{amsfonts}
\usepackage{amssymb}
\usepackage{epsf}
\usepackage{cite}
\usepackage{balance}
\usepackage{fancyhdr}
\usepackage{graphicx}
\usepackage{subfigure}
\usepackage[blocks]{authblk}
\usepackage[stable]{footmisc}
\usepackage{float}
\usepackage{fancyhdr}
\usepackage{gensymb}
\usepackage[bookmarks=false]{hyperref}
\usepackage{epstopdf}

\makeatletter
\def\ps@IEEEtitlepagestyle{%
  \def\@oddfoot{\mycopyrightnotice}%
  \def\@evenfoot{}%
}
\def\mycopyrightnotice{%
  {\footnotesize \copyright 2015 IEEE. Personal use of this material is permitted. Permission from IEEE must be obtained for all other uses, in any current or future media\hfill}% <--- Change here
  \gdef\mycopyrightnotice{}% just in case
}

\newcounter{subeq}

\IEEEoverridecommandlockouts
\date{}
\begin{document}
\title{Indoor Location Estimation with Optical-based OFDM Communications}

\author{Mohammadreza~Aminikashani,~\IEEEmembership{Student Member,~IEEE,}
        Wenjun Gu,~\IEEEmembership{Student Member,~IEEE,}
        and~Mohsen~Kavehrad,~\IEEEmembership{Fellow,~IEEE}

\thanks{Mohammadreza Aminikashani, Wenjun Gu and Mohsen Kavehrad are with the Pennsylvania State University, University Park, PA 16802 (email:  {amini, wzg112,  mkavehrad}@psu.edu).}}

\maketitle
%\balance
\begin{abstract}
Visible Light Communication (VLC) using light emitting diodes (LEDs) has been gaining increasing attention in recent years as it is appealing for a wide range of applications such as indoor positioning. Orthogonal frequency division multiplexing (OFDM) has been applied to indoor wireless optical communications in order to mitigate the effect of multipath distortion of the optical channel as well as increasing data rate. In this paper, a novel OFDM VLC system is proposed which can be utilized for both communications and indoor positioning. A positioning algorithm based on power attenuation is used to estimate the receiver coordinates. We further calculate the positioning errors in all the locations of a room and compare them with those using single carrier modulation scheme, i.e., on-off keying (OOK) modulation. We demonstrate that OFDM positioning system outperforms its conventional counterpart. Finally, we investigate the impact of different system parameters on the positioning accuracy of the proposed OFDM VLC system.
 \end{abstract}

\section{Introduction}\label{INTRODUCTION}
\IEEEPARstart{V}{isible} light communication has been extensively studied recently as a promising complementary and/or alternative technology to its radio frequency (RF) counterparts particularly in indoor environments. VLC systems offer many attractive features such as higher bandwidth capacity, robustness to electromagnetic interference, excellent security and low cost deployment \cite{1.1,1.2,1.3,peng,1.4,1}. These systems however suffer from multipath distortion due to dispersion of the optical signal caused by reflections from various sources inside a room.

Orthogonal frequency division multiplexing has been proposed in the literature to combat intersymbol interference (ISI) caused by multipath reflections \cite{shieh2008coherent,gonzalez2005ofdm,elgala2009indoor,armstrong2009ofdm,kahn1997wireless,armstrong2006power}. OFDM is capable of employing very low-complexity equalization with single-tap equalizers in the frequency domain, and allows adaptive modulation and power allocation. There have been several OFDM techniques for VLC systems using intensity-modulation direct-detection (IM/DD) including DC-clipped OFDM \cite{kahn1997wireless}, asymmetrically clipped optical OFDM (ACO-OFDM) \cite{armstrong2006power}, PAM-modulated discrete multitone (PAM-DMT) \cite{lee2009pam}, Flip-OFDM \cite{flip} and unipolar OFDM (U-OFDM) \cite{UOFDM}. In DC-clipped OFDM, a DC bias is added to the signal to make it unipolar and suitable for optical transmission. Hard-clipping on the negative signal peaks is used in order to reduce the DC bias required. The other techniques have been proposed to remove the biasing requirement and therefore improve the energy efficiency of DC-clipped OFDM. Particularly, ACO-OFDM and PAM-DMT clip the entire negative excursion of the waveform. In ACO-OFDM, to avoid the impairment from clipping noise, only odd subcarriers are modulated by information symbols. In PAM-DMT, only the imaginary parts of the subcarriers are modulated such that clipping noise falls only on the real part of each subcarrier and becomes orthogonal to the desired signal. On the other hand, U-OFDM and Flip-OFDM extract the negative and positive samples from the real bipolar OFDM symbol and separately transmit these two components over two successive OFDM frames where the polarity of the negative samples is inverted before transmission. As discussed in \cite{awgn}, all of these four non-biasing OFDM approaches exhibit the same performance in an additive white Gaussian noise (AWGN) channel.

In current indoor visible light positioning systems, several algorithms have been proposed to calculate the receiver coordinates. In one approach, a photo diode (PD) is employed to detect received signal strength (RSS) information. The distancse between transmitter and receiver is then estimated based on the power attenuation, and the receiver coordinates are calculated by lateration algorithm \cite{5,6}. In another approach, RSS information is pre-detected by a PD for each location and stored as fingerprint in the offline stage. By matching the stored fingerprints with the RSS feature of the current location, the receiver location is estimated in the online stage \cite{3}. In \cite{4}, proximity positioning concept has been used relying on a grid of transmitters as reference points, each of which has a known coordinate. The mobile receiver is assigned the same coordinates as the reference point sending the strongest signal. Image sensor is another form of receiver which detects angle of arrival (AOA) information for the angulation algorithm used to calculate the receiver location \cite{2}. Other techniques have been also proposed for VLC systems to improve the indoor positioning performance. In \cite{7}, Gaussian mixture sigma point particle filter technique has been applied to increase the accuracy of the estimated coordinates. Accelerometer has been employed in \cite{acc} such that the information on the receiver height is not required. To the best of our knowledge, the previous studies have been built on the assumption of a low speed single carrier modulation or/and have not considered the multipath reflections. However, a practical VLC system would be likely to deploy the same configuration for both positioning and communication purposes where high speed data rates are desired. Furthermore, it has been shown in \cite{gu2015,gu2015impact,aminikashani2015indoor} that multipath reflections can severely degrade the positioning accuracy especially in the corner and the edge areas of a room.

In this paper, to mitigate the multipath reflections as well as providing a high data rate transmission, we propose an OFDM VLC system that can be used for both indoor positioning and communications. The positioning algorithm employed is based on RSS information detected by a PD and the lateration technique. We show that our proposed system can achieve an excellent accuracy even in dispersive optical channels and for very low signal power values.

The rest of the paper is organized as follows. In Section II, the system model and OFDM system configuration are briefly introduced. In Section III, the positioning algorithm is described. In Section IV, we present numerical results on the positioning accuracy of the proposed OFDM VLC system and compare its performance with that of its OOK counterpart. The effect of different system parameters on the positioning accuracy is also investigated. Finally, Section V concludes the paper.

\section{System Configuration}
\subsection{System Model}
\begin{figure}
\centering
\includegraphics[width = 8cm, height = 7.5cm]{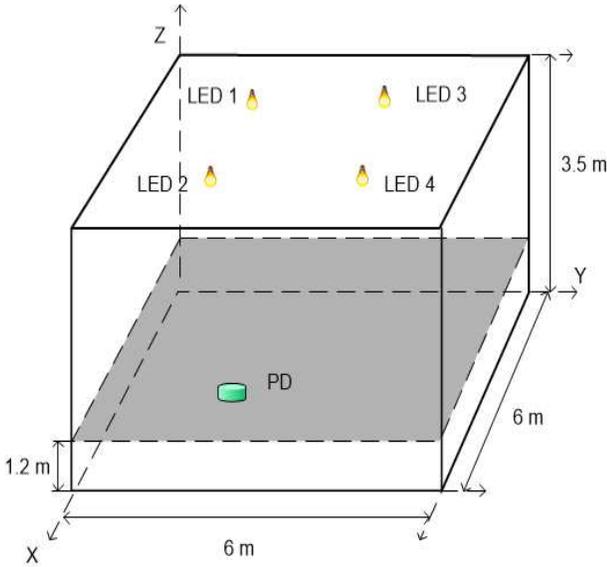}
\caption{System Configuration.}
\end{figure}
We consider a typical room shown in Fig. 1 with dimensions of 6 m $\times$ 6 m $\times$ 3.5 m where four LED bulbs are located at height of 3.3 m with a rectangular layout. Data are transmitted from these LED bulbs after they are modulated by driver circuits. Each LED bulb has an identification (ID) denoting its coordinates which is included in the transmitted data. A PD as the receiver is located at the height of 1.2 m and has a field of view (FOV) of 70$^{\circ}$ and a receiving area of 1 cm$^2$. Strict time domain multiplexing is used where the entire OFDM frequency spectrum is assigned to a single LED transmitter for at least one OFDM symbol including a cyclic prefix (CP).
\subsection{Optical Wireless Channel}
\begin{table}
\begin{center}
\begin{quote}
\caption{SYSTEM PARAMETERS}
\label{table2}
\end{quote}
%\resizebox{\columnwidth}{!}{
\begin{tabular}{|c|c|}
  \hline
  \textbf{Room dimensions} &\textbf{Reflection coefficients} \\ \hline
  length: 6 m &    ${{\rho }_{wall}}$: 0.66\\
  width: 6 m  & ${{\rho }_{Ceiling}}$: 0.35\\
  height: 3.5 m & ${{\rho }_{Floor}}$: 0.60\\ \hline
 \textbf{Transmitters (Sources)} &\textbf{Receiver} \\ \hline
  Wavelength: 420 nm &Area $\left(A\right)$: 1$\times 10^{-4} \text{m}^{2}$\\
  Height $\left(H\right)$: 3.3 m &Height $\left(h\right)$: 1.2 m\\
 Lambertian mode $\left(m\right)$: 1 &Elevation: +90$^{\circ}$\\
 Elevation: -90$^{\circ}$ &Azimuth: 0$^{\circ}$\\
 Azimuth: 0$^{\circ}$ &FOV $\left({{\Psi }_{c}}\right)$: 70$^{\circ}$\\
 Coordinates: (2,2) (2,4) (4,2) (4,4) & \\
 Power for "1"/ "0": 5 W/3 W & \\ \hline
\end{tabular}
\end{center}
\end{table}
We assume an indoor optical multipath channel where transmitters and a receiver are placed in a room whose configuration is summarized in Table I. The baseband channel model including noise is expressed as
\begin{equation}\label{1ch}
y\left( t \right)=\eta x\left( t \right)*h\left( t \right)+\omega \left( t \right)
\end{equation}
where $y\left( t \right)$ is the received electrical signal, $\eta $ is the photodetector responsivity, $h\left( t \right)$ is the multipath impulse response of the optical channel, and $\omega \left( t \right)$ denotes ambient light shot noise and thermal noise.

Combined deterministic and modified Monte Carlo (CDMMC) method recently developed by Chowdhury \emph{et al} \cite{12} is used to simulate the impulse response of the optical wireless channel. Deterministic approaches \cite{12.1,12.2} proposed to approximate impulse response of indoor optical wireless channels divide the reflecting surfaces into small elements and provide the best accuracy, but at the cost of high computing time. On the other hand, modified Monte Carlo (MMC) approaches \cite{12.3,12.4} calculate the impulse responses very fast, but the calculated impulse responses are not as temporally smooth when compared to deterministic approaches. The algorithm in \cite{12} takes advantage of both deterministic and modified Monte Carlo methods. In particular, the contribution of the first reflections to the total impulse response is calculated by a deterministic method for high accuracy, while an MMC method is employed to calculate the second and higher order reflections and achieve a lower execution time.

We consider the LOS and first three reflections to simulate the impulse response of the channel. For each transmitter, we generate 64 different channels by placing the receiver in different locations within the room with the same height (i.e., 1.2 m). Figs. 2-4 demonstrate the contributions from different orders of reflections to the total impulse responses for three exemplary locations inside the room representing weak to strong scatterings and multipath reflections.
\begin{figure}
\centering
\includegraphics[width = 8cm, height = 7.5cm]{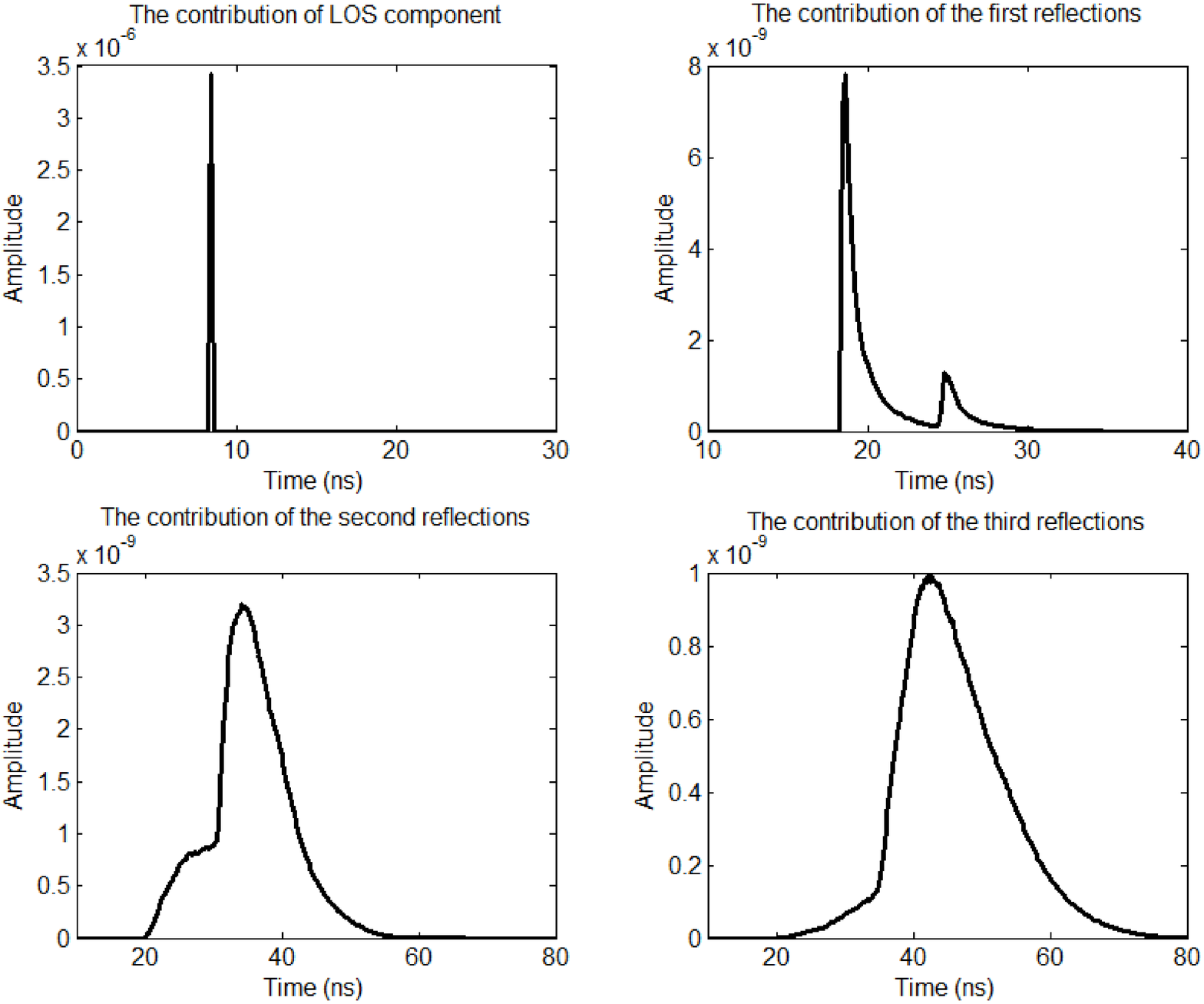}
\caption{The contributions from different orders of reflections to the total impulse response of a location at the center of the room (weak scatterings and multipath reflections).}
\end{figure}
\begin{figure}
\centering
\includegraphics[width = 8cm, height = 7.5cm]{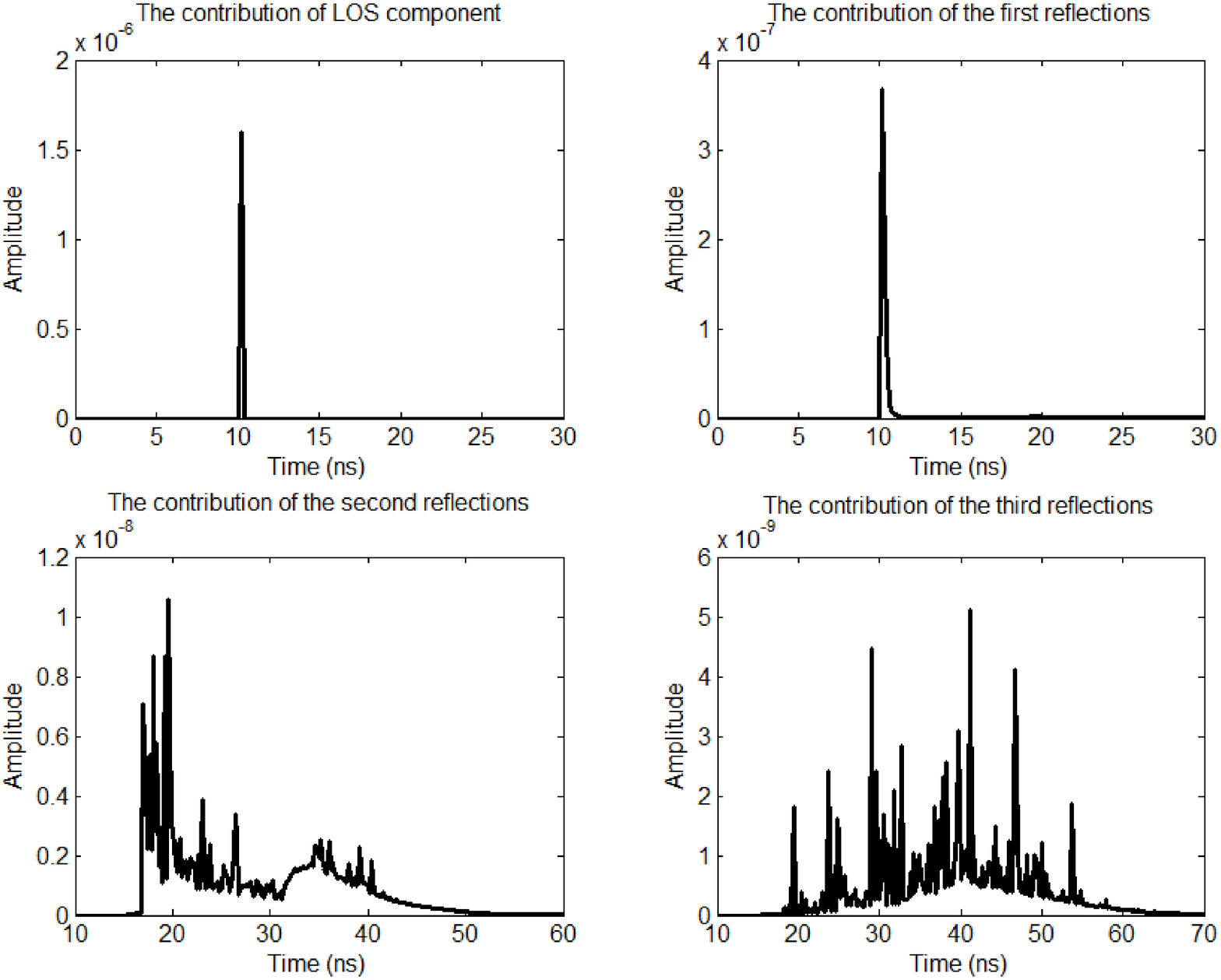}
\caption{The contributions from different orders of reflections to the total impulse response of a location at the edge of the room (medium scatterings and multipath reflections).}
\end{figure}
\begin{figure}
\centering
\includegraphics[width = 8cm, height = 7.5cm]{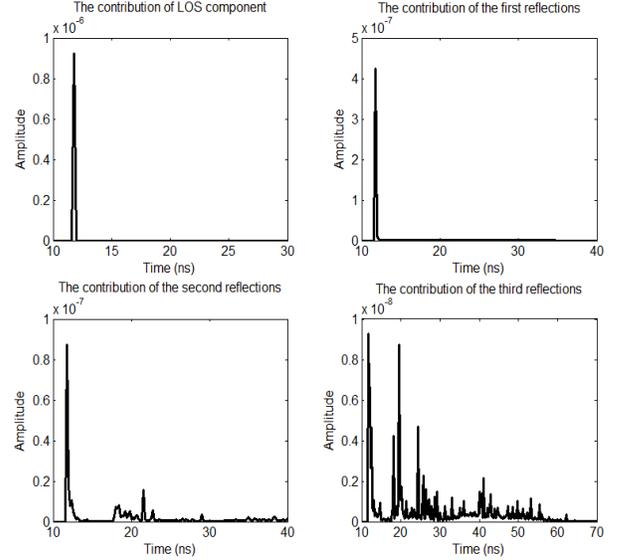}
\caption{The contributions from different orders of reflections to the total impulse response of a location at the corner of the room (strong scatterings and multipath reflections).}
\end{figure}

\subsection{OFDM transmitter and receiver}
Different OFDM techniques have been proposed for optical wireless communications in the literature. For the sake of brevity, ACO-OFDM is only considered in this paper as it utilizes a large dynamic range of LED and thus is more efficient in terms of optical power than systems using DC-biasing. However, the generalization to other techniques is very straightforward. A block diagram of an ACO-OFDM communication and positioning system is depicted in Fig. 5. The data and LED ID code are combined as the input bits which is parsed into a set of $N/4$ complex data symbols denoted by $\mathbf{I}={{\left[ {{I}_{0}},{{I}_{1}},...{{I}_{{N}/{4}\;-1}} \right]}^{T}}$ where $N$ is the number of subcarriers, and ${{\left( . \right)}^{T}}$ indicates the transpose of a vector. These symbols are drawn from constellations such as $M$-QAM or $M$-PSK where $M$ is the constellation size. For VLC systems using IM/DD, a real valued signal is required to modulate the LED intensity. Thus, ACO-OFDM subcarriers must have Hermitian symmetry. In ACO-OFDM, impairment from clipping noise is avoided by mapping the complex input symbols onto an $N\times 1$ vector $\mathbf{S}$ as
\begin{equation}
\mathbf{S}={{\left[ 0,{{I}_{0}},0,{{I}_{1}},...,0,{{I}_{N-1}},0,I_{N-1}^{*},0,...,I_{1}^{*},0,I_{0}^{*},0 \right]}^{T}}
\end{equation}
where ${{\left( . \right)}^{*}}$ denotes the complex conjugate of a vector. An $N$-point inverse fast Fourier transform (IFFT) is then applied creating the time domain signal $\mathbf{x}$. A CP is added to the real valued output signal turning the linear convolution with the channel into a circular one to mitigate inter-carrier interference (ICI) as well as inter-block interference (IBI). All the negative values of the transmitted signal are clipped to zero to make it unipolar and suitable for optical transmission. This clipping operation does not affect the data-carrying subcarriers but decreases their amplitude to exactly a half. The clipped signal is then converted to analog and finally modulates the intensity of an LED.

At the receiver, the signal is detected by a PD and then converted back to a digital signal. The CP is removed and an $N$-point fast Fourier transform (FFT) is applied on the electrical OFDM signal. The training sequence is employed for synchronization and channel estimation as discussed in \cite{bilal}. A single tap equalizer is then used for each subcarrier to compensate for channel distortion and the transmitted symbols are recovered from the odd subcarriers and denoted by $\mathbf{\hat{I}}={{\left[ {{{\hat{I}}}_{0}},{{{\hat{I}}}_{1}},...{{{\hat{I}}}_{{N}/{4}\;-1}} \right]}^{T}}$. The LED ID is decoded and the transmitter coordinates are obtained which are fed to the positioning block along with the estimated channel DC gain as shown in Fig. 5. The receiver coordinates are finally estimated by employing the positioning algorithm detailed in the following section.
\begin{figure*}[t]
\centering
\includegraphics[width = 13cm, height = 6cm]{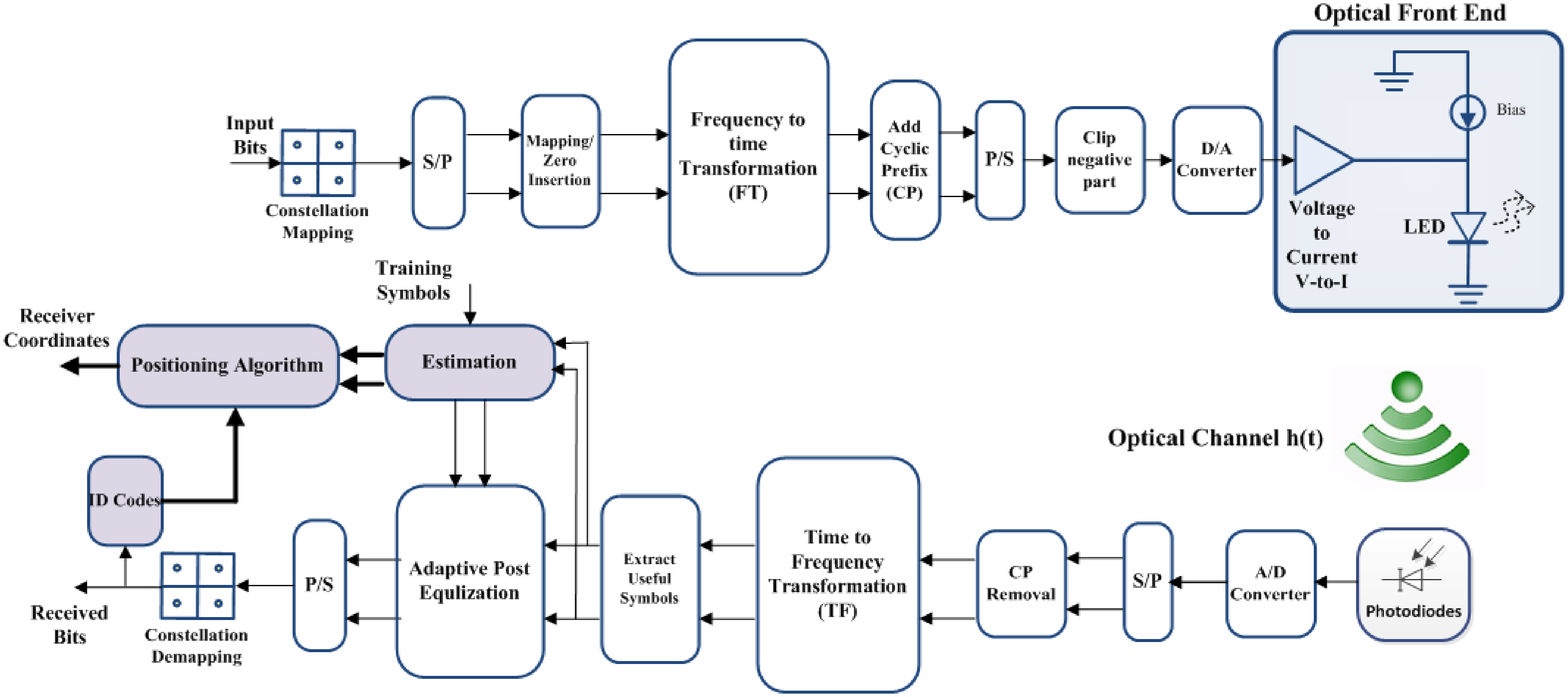}
\caption{OFDM transmitter and receiver configuration for both positioning and communication purposes.}
\end{figure*}
\section{POSITIONING ALGORITHM}
For the system under consideration, the received optical power from the $k^{\text{th}}$ transmitter, $k=1,2,3$ and $4$, can be expressed as
\begin{equation}\label{eq1.1}
{{P}_{r,k}}={{H}_{k}}(0){{P}_{t,k}}
\end{equation}
where ${{P}_{t}}$ denotes the transmitted optical power from the $k^{\text{th}}$ LED bulb, and ${{H}_{k}}\left( 0 \right)$ is the channel DC gain that can be obtained as \cite{zhang2014}
\begin{equation}\label{eq1.2}
{{H}_{k}}\left( 0 \right)=\frac{m+1}{2\pi d_{k}^{2}}A {{\cos }^{m}}({{\phi }_{k}}){{T}_{s}}({{\psi }_{k}})g({{\psi }_{k}})\cos ({{\psi }_{k}})
\end{equation}
In (\ref{eq1.2}), $A$ is the physical area of the detector, ${{\psi }_{k}}$ is the angle of incidence with respect to the receiver axis, ${{T}_{s}}\left( {{\psi }_{k}} \right)$ is the gain of optical filter, $g\left( {{\psi }_{k}} \right)$ is the concentrator gain, ${{\phi }_{k}}$ is the angle of irradiance with respect to the transmitter perpendicular axis, $d_k$ is the distance between transmitter $k$ and receiver, and $m$ is the Lambertian order. Assuming that both receiver and transmitter axes are perpendicular to the ceiling, ${{\phi }_{k}}$ and ${{\psi }_{k}}$ are equal and can be estimated as
\begin{equation}\label{eq2}
\cos \left( {{\psi }_{k}} \right)=\cos \left( {{\phi }_{k}} \right)=\left( H-h \right)/{{d}_{k}}
\end{equation}
where $H$ and $h$ are the transmitter and receiver heights, respectively. For a compound parabolic concentrator (CPC), $g\left( {{\psi }_{k}} \right)$ is defined as
\begin{equation}\label{eq5}
g\left( \psi_{k}  \right)=\begin{cases}
\frac{{{n}^{2}}}{{{\sin }^{2}}\left( {{\Psi }_{c}} \right)}, & 0\le \psi_{k} \le {{\Psi }_{c}}  \\
0, & \psi_{k} >{{\Psi }_{c}}  \\
\end{cases}
\end{equation}
where $n$ and ${{\Psi }_{c}}$ respectively denote the refractive index and the FOV of the concentrator.

For the proposed OFDM system, the channel DC gain can be well estimated as
\begin{equation}\label{eq3}
{{\tilde{H}}_{k}}\left( 0 \right)=\frac{4}{N}\sum\limits_{i=1}^{{N}/{4}}{{{P}_{k,i}}}
\end{equation} 	
where ${{P}_{k,i}}$ is the power attenuation of the ${{i}^{th}}$ symbol transmitted from the ${{k}^{th}}$ transmitter and is obtained using the training symbols used for synchronization as
\begin{equation}\label{eq4}
{{P}_{k,i}}={{\left| \frac{{{{\hat{I}}}_{k,i}}}{{{I}_{k,i}}} \right|}}.
\end{equation}

Considering (\ref{eq1.1})-(\ref{eq4}), $d_k$ can be calculated as
\begin{equation}\label{eq7}
d_{k}^{m+3}=\frac{\left( m+1 \right)A{{T}_{s}}\left( {{\psi }_{k}} \right)g\left( {{\psi }_{k}} \right){{\left( H-h \right)}^{m+1}}}{2\pi {{\tilde{H}}_{k}}}.
\end{equation}
Horizontal distance between the ${{k}^{th}}$ transmitter and the receiver can be estimated as
\begin{equation}\label{eq8}
{{r}_{k}}=\sqrt{{{d}_{k}}^{2}-{{\left( H-h \right)}^{2}}}.
\end{equation}
Then, according to the lateration algorithm \cite{14,30}, a set of four quadratic equations can be formed as follows
\begin{equation}\label{eq9}
 \begin{cases}
  {{\left( x-{{x}_{1}} \right)}^{2}}+{{\left( y-{{y}_{1}} \right)}^{2}}=r_{1}^{2}& \\
 {{\left( x-{{x}_{2}} \right)}^{2}}+{{\left( y-{{y}_{2}} \right)}^{2}}=r_{2}^{2}& \\
 {{\left( x-{{x}_{3}} \right)}^{2}}+{{\left( y-{{y}_{3}} \right)}^{2}}=r_{3}^{2}& \\
 {{\left( x-{{x}_{4}} \right)}^{2}}+{{\left( y-{{y}_{4}} \right)}^{2}}=r_{4}^{2}& \\
\end{cases},
\end{equation}
where $\left( x,y \right)$ is the receiver coordinates to be estimated and $\left( {{x}_{k}},{{y}_{k}} \right)$ is the ${{k}^{th}}$ transmitter coordinates obtained from the recovered LED ID in a two-dimensional space.

By subtracting the first equation from the last three equations, we obtain
\begin{align}\nonumber
&\left( {{x}_{1}}-{{x}_{j}} \right)x+\left( {{y}_{1}}-{{y}_{j}} \right)y=\\\label{eq10}
&\left( r_{j}^{2}-r_{1}^{2}-x_{j}^{2}+x_{1}^{2}-y_{j}^{2}+y_{1}^{2} \right)/2
\end{align}
where $j=2,3$ and $4$. (\ref{eq10}) can be formed in a matrix format as $\mathbf{AX}=\mathbf{B}$ where $\mathbf{A}$, $\mathbf{B}$ and $\mathbf{X}$ are defined as
\begin{equation}
\mathbf{A}=\left[ \begin{matrix}
   {{x}_{2}}-{{x}_{1}} & {{y}_{2}}-{{y}_{1}}  \\
   {{x}_{3}}-{{x}_{1}} & {{y}_{3}}-{{y}_{1}}  \\
   {{x}_{4}}-{{x}_{1}} & {{y}_{4}}-{{y}_{1}}  \\
\end{matrix} \right],
\end{equation}
\begin{equation}
\mathbf{B}=\frac{1}{2}\left[ \begin{matrix}
   \left( r_{1}^{2}-r_{2}^{2} \right)+\left( x_{2}^{2}+y_{2}^{2} \right)-\left( x_{1}^{2}+y_{1}^{2} \right)  \\
   \left( r_{1}^{2}-r_{3}^{2} \right)+\left( x_{3}^{2}+y_{3}^{2} \right)-\left( x_{1}^{2}+y_{1}^{2} \right)  \\
   \left( r_{1}^{2}-r_{4}^{2} \right)+\left( x_{4}^{2}+y_{4}^{2} \right)-\left( x_{1}^{2}+y_{1}^{2} \right)  \\
\end{matrix} \right],
\end{equation}
\begin{equation}
\mathbf{X}={{[x\ y]}^{T}}.
\end{equation}
The estimated receiver coordinates can then be obtained by the linear least squares estimation approach as \cite{14}
\begin{equation}
\mathbf{\hat{X}}={{({{\mathbf{A}}^{\mathbf{T}}}\mathbf{A})}^{-1}}{{\mathbf{A}}^{\mathbf{T}}}\mathbf{B}.
\end{equation}
\section{SIMULATION AND ANALYSIS}
In this section, we present numerical results for the proposed indoor VLC system. In the following, we consider an OFDM system with a number of subcarriers of $N=64$, $256$, $512$ or $1024$ where the symbols are drawn from an $M$-QAM modulation constellation. We set the CP length three times of the root mean square (RMS) delay spread of the worst impulse response and assume a data with minimum rate of 25 Mbps \footnote{The data rate is 25 Mbps for 4-QAM modulation scheme. For higher-order modulations (i.e., $M>4$), the bit rate can be achieved as $R =$ 25 Mbps$\times {{\log }_{2}}\left( M \right)$.}. The sum of ambient light shot noise and receiver thermal noise is modeled as real baseband AWGN with zero mean and power of -10 dBm \cite{o2007optical,grubor2008bandwidth}. Furthermore, to take LED nonlinearity into account, OPTEK, OVSPxBCR4 1-Watt white LED is considered in simulations whose optical and electrical characteristics are given in Table II. A polynomial order of five is used to realistically model the measured transfer function. Fig. 6 demonstrates the non-linear transfer characteristics of the LED from the data sheet and using the polynomial function. The four OPTEK LEDs are biased at 3.2V.

\begin{table}
\begin{center}
\begin{quote}
\caption{Optical and Electrical Characteristics of OPTEK, OVSPxBCR4 1-Watt white LED.}
\label{table2}
\end{quote}
{\small{
\begin{tabular}{|c|c|c|c|c|c|}
  \hline
  \textbf{Symbol} &\textbf{Parameter}  &\textbf{MIN} &\textbf{TYP} &\textbf{MAX} &\textbf{Units}\\ \hline
  $V_F$  &Forward Voltage &3.0 &3.5 &4 &$V$\\ \hline
  $\Phi$  &Luminous Flux &67 &90 &113 &lm\\ \hline
  ${{\Theta }^{1/2}}$  &50\% Power Angle &--- &120 &--- &deg\\ \hline
  \end{tabular}}}
\end{center}
\end{table}
\begin{figure} \centering
\subfigure[]{\includegraphics[width = 7cm, height = 5cm]{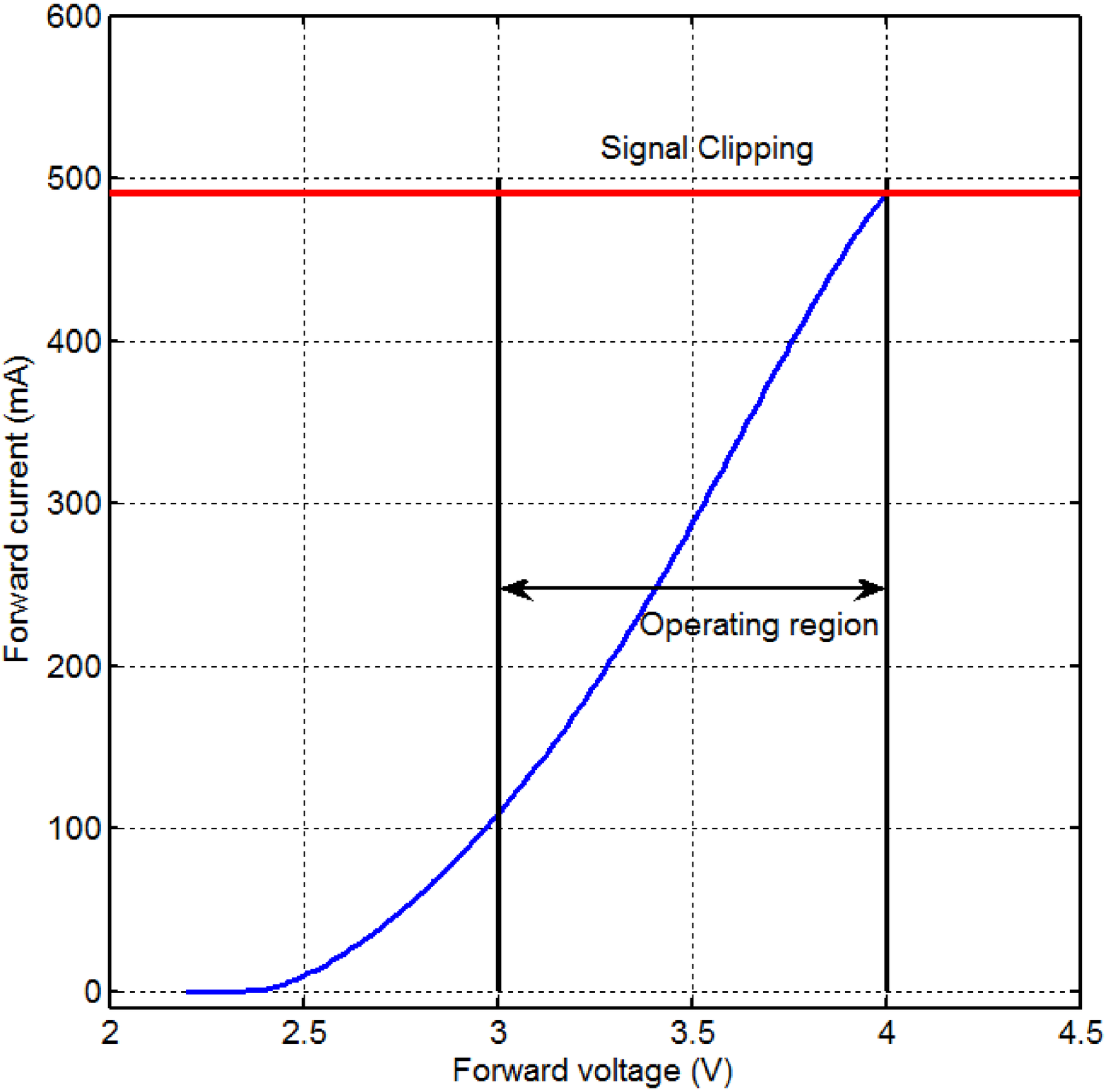}}
\subfigure[]{\includegraphics[width = 7cm, height = 5cm]{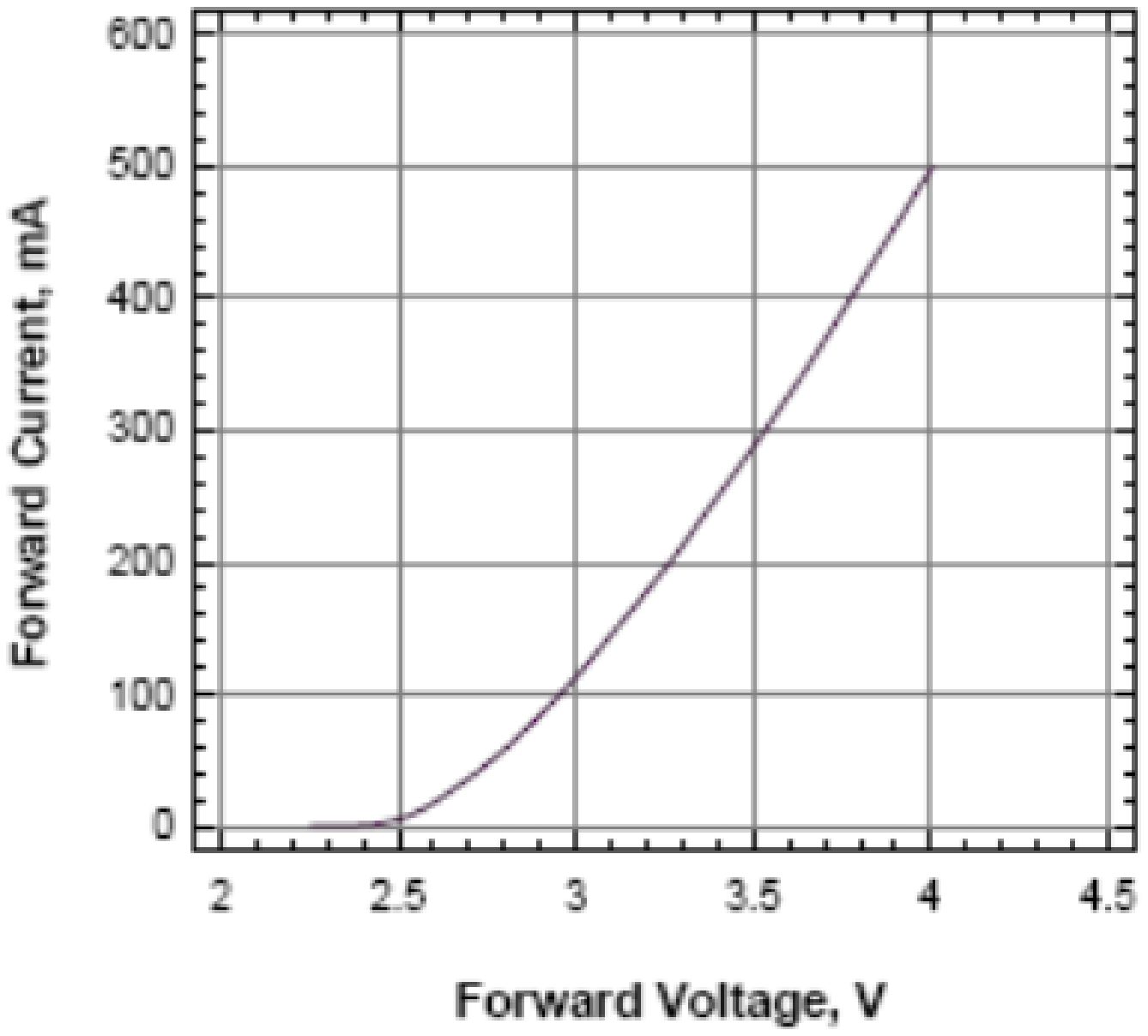}}
\caption{Transfer characteristics of OPTEK, OVSPxBCR4 1-Watt white LED. (a) Fifth-order polynomial fit to the data. (b) The curve from the data sheet.}
\label{led}
\end{figure}
\subsection{Performance comparison of single- and multi-carrier modulation schemes}
In this subsection, the positioning performance of the proposed OFDM system is compared with the performance of those using single carrier modulation scheme, i.e., OOK. We assume that the average electrical power of the transmitted signal before modulating each LED is $P_{t_e,k}=$5 dBm. Thus, the simulated electrical signal-to-noise ratio (SNR) is 15 dB.

Fig. 7 demonstrates the positioning error distribution over the room for an indoor OFDM VLC system with 4-QAM modulation and the FFT size of 512. As it can be seen, the positioning errors are very small for the most locations inside the room, but become larger when the receiver approaches the corners and edges due to the severity of the multipath reflections.
\begin{figure}
\centering
\includegraphics[width = 8cm, height = 7cm]{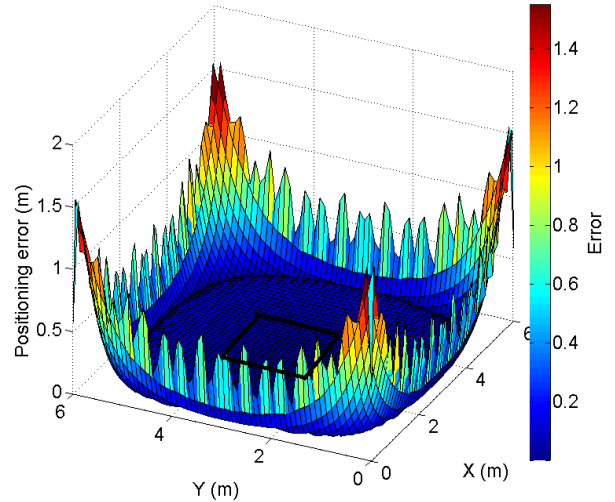}
\caption{Positioning error distribution for OFDM system with 4-QAM modulation, $N$ = 512 and $P_{t_e,k}$ = 5 dBm.}
\end{figure}

Fig. 8, on the other hand, shows the positioning error distribution over the room for an indoor VLC system employing OOK modulation with the same data rate as that of the OFDM system with 4-QAM modulation (i.e., 25 Mbps). As observed, the positioning errors are relatively small within the rectangle shown in Fig. 8 where the LED bulbs are located right above its corners. However, the positioning error becomes significantly larger when the receiver moves toward the corners and edges as the effect of the multipath reflections increases.
\begin{figure}
\centering
\includegraphics[width = 7cm, height = 6.5cm]{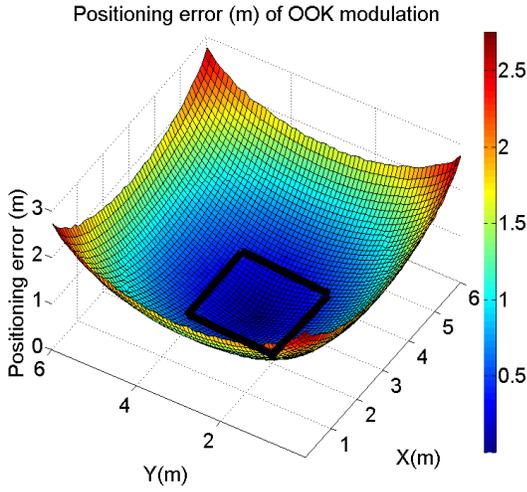}
\caption{Positioning error distribution for OOK modulation with $P_{t_e,k}$ = 5 dBm.}
\end{figure}

Figs. 9 and 10 present the histograms of the positioning errors for OFDM and OOK modulation schemes, respectively. For OFDM modulation, most of the positioning errors are less than 0.1 m and only a few of them are more than 1 m corresponding to the corner area. However, for OOK modulation, the positioning errors are widely spread from zero to around 2.3 m, and only a few of them are less than 0.1 m that correspond to the central area. From Figs. 7-10, it can be clearly seen that the OFDM system outperforms its OOK counterpart.
\begin{figure}
\centering
\includegraphics[width = 7cm, height = 6cm]{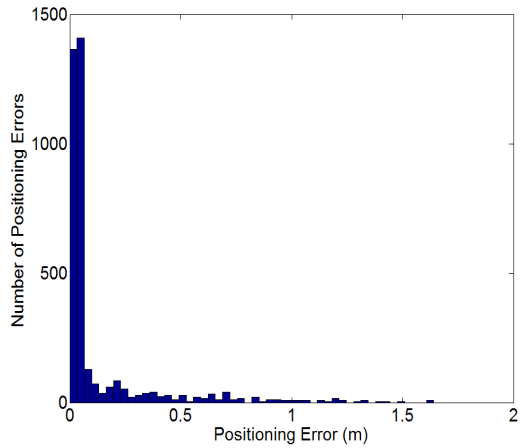}
\caption{Histogram of positioning errors for OFDM system with 4-QAM modulation, $N$ = 512 and $P_{t_e,k}$ = 5 dBm.}
\end{figure}
\begin{figure}
\centering
\includegraphics[width = 7cm, height = 6cm]{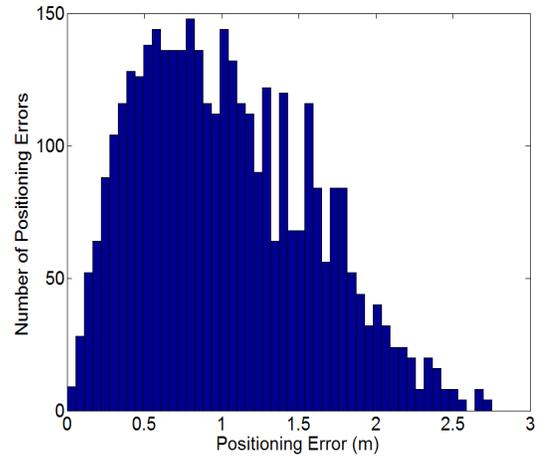}
\caption{Histogram of positioning errors for OOK modulation with $P_{t_e,k}$ = 5 dBm.}
\end{figure}

Table III summarizes and compares the positioning errors of OFDM and OOK modulation schemes. As seen, OFDM modulation provides a much better positioning accuracy than OOK modulation for all the locations inside the room. Particularly, the RMS error is 0.08 m for the rectangular area covered perfectly by the four LED bulbs when OFDM modulation is used while it is 0.43 m for OOK modulation. The total RMS errors are 0.2609 m and 1.01 m for OFDM and OOK modulation schemes as the rectangular area covered by LED bulbs is only 11.1\% of the total area. It should be noted that the average positioning accuracy can be increased by optimizing the layout design of the LED bulbs in future.
\begin{table}
\begin{center}
\begin{quote}
\caption{Positioning error for single- and multi-carrier modulation schemes}
\label{table2}
\end{quote}
\begin{tabular}{|c|c|c|}
  \hline
  Positioning error (m) &\textbf{OFDM}   &\textbf{OOK}\\
  &\textbf{modulation (m)} & \textbf{modulation (m)}\\ \hline
  Corner (0, 0)  &0.578 &2.18 \\ \hline
  Edge (3 m, 0)  &0.49 &1.53 \\ \hline
  Center (3 m, 3 m)  &2$\times10^{-6}$ &10$^{-5}$ \\ \hline
  RMS error of  &0.08 &0.43 \\
  the rectangular area & &\\\hline
  RMS error of   &0.2609 &1.01 \\
  the whole room & & \\\hline
  \end{tabular}
\end{center}
\end{table}
\subsection{Effect of signal power on the positioning accuracy}
In this subsection, we investigate the effect of the average electrical power of the transmitted signal on the positioning accuracy of the proposed OFDM VLC system. We consider an OFDM system with  4-QAM modulation and $N=512$. Figs. 11 and 12 present the positioning error distribution for the OFDM system with electrical SNR of 0 dB and 30 dB. The total RMS errors are calculated as 0.384 m and 0.2766 m for SNR values of 0 dB and 30 dB, respectively.
\begin{figure}
\centering
\includegraphics[width = 7cm, height = 6.5cm]{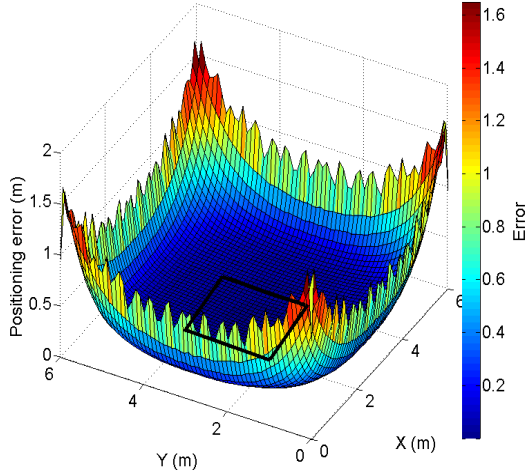}
\caption{Positioning error distribution for OFDM system with 4-QAM modulation, $N$ = 512 and $P_{t_e,k}$ = -10 dBm.}
\end{figure}
\begin{figure}
\centering
\includegraphics[width = 7cm, height = 6.5cm]{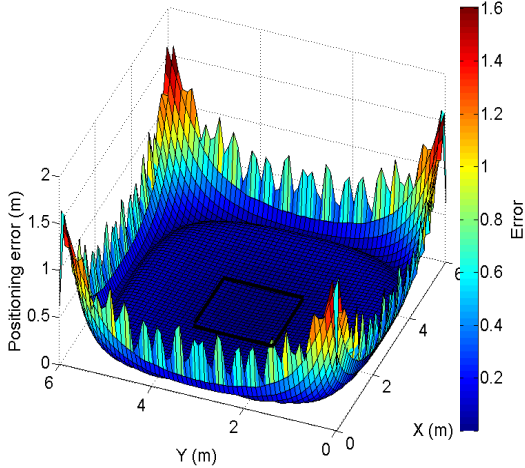}
\caption{Positioning error distribution for OFDM system with 4-QAM modulation, $N$ = 512 and $P_{t_e,k}$ = 20 dBm.}
\end{figure}

Figs. 13 and 14 further demonstrate the corresponding histograms of the positioning errors for different SNR values. It is apparent from Figs. 11 and 13 that our proposed positioning system works satisfactorily even at very low SNR values resulting from dimming and shadowing effects. Furthermore, according to Figs. 7, 9 and 11-14 and as expected, increasing the average electrical power of the transmitted signal results in a better performance. However, at very high power values, nonlinearity distortion effects dominate the performance and the positioning accuracy decreases. It is the main reason the performance of the VLC system with 30 dB SNR is slightly worse than that of 15 dB SNR presented earlier. Therefore, for an OFDM indoor VLC positioning system, there is an optimum power value that depends on the LED characteristics.
\begin{figure}
\centering
\includegraphics[width = 7cm, height =6cm]{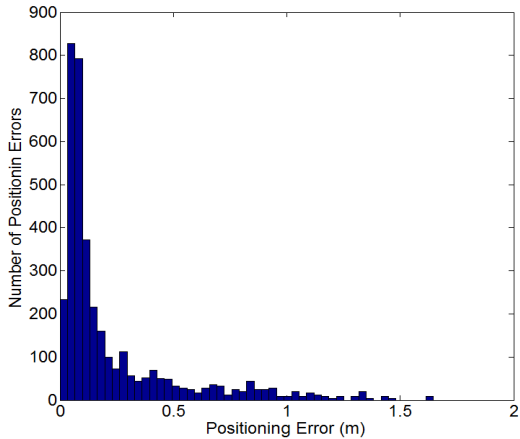}
\caption{Histogram of positioning errors for OFDM system with 4-QAM modulation, $N$ = 512 and $P_{t_e,k}$ = -10 dBm.}
\end{figure}
\begin{figure}
\centering
\includegraphics[width = 7cm, height = 6cm]{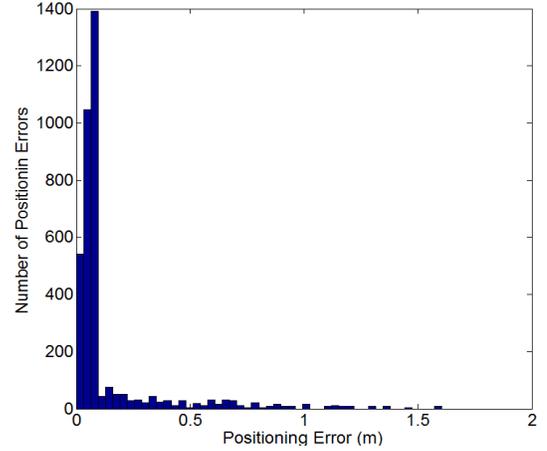}
\caption{Histogram of positioning errors for OFDM system with 4-QAM modulation, $N$ = 512 and $P_{t_e,k}$ = 20 dBm.}
\end{figure}
\subsection{Effect of modulation order on the positioning accuracy}
Here, we analyze the impact of the modulation order on the positioning performance. Figs. 15 and 16 show the positioning error distribution of the OFDM system with the FFT size of 512 and 15 dB SNR employing 16- and 64-QAM modulation, respectively. The corresponding histograms of the positioning errors are shown in Figs. 17 and 18. The total RMS error is obtained as 0.2665 m for 16-QAM and 0.2716 m for 64-QAM. By comparing these RMS error values with the one calculated for 4-QAM modulation, we observe that all three systems yield nearly the same performance. Thus, the constellation size does not have a significant effect on the positioning performance of the proposed OFDM VLC system. It clearly shows that our proposed channel DC gain estimation works perfectly for high-order constellations as well.

\begin{figure}
\centering
\includegraphics[width = 7cm, height = 6.5cm]{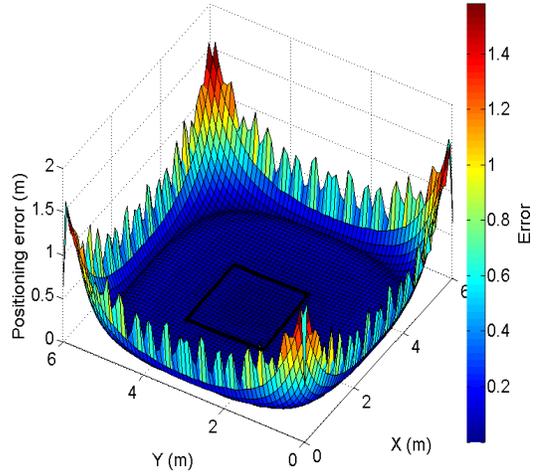}
\caption{Positioning error distribution for OFDM system with 16-QAM modulation, $N$ = 512 and $P_{t_e,k}$ = 5 dBm.}
\end{figure}
\begin{figure}
\centering
\includegraphics[width = 7cm, height = 6.5cm]{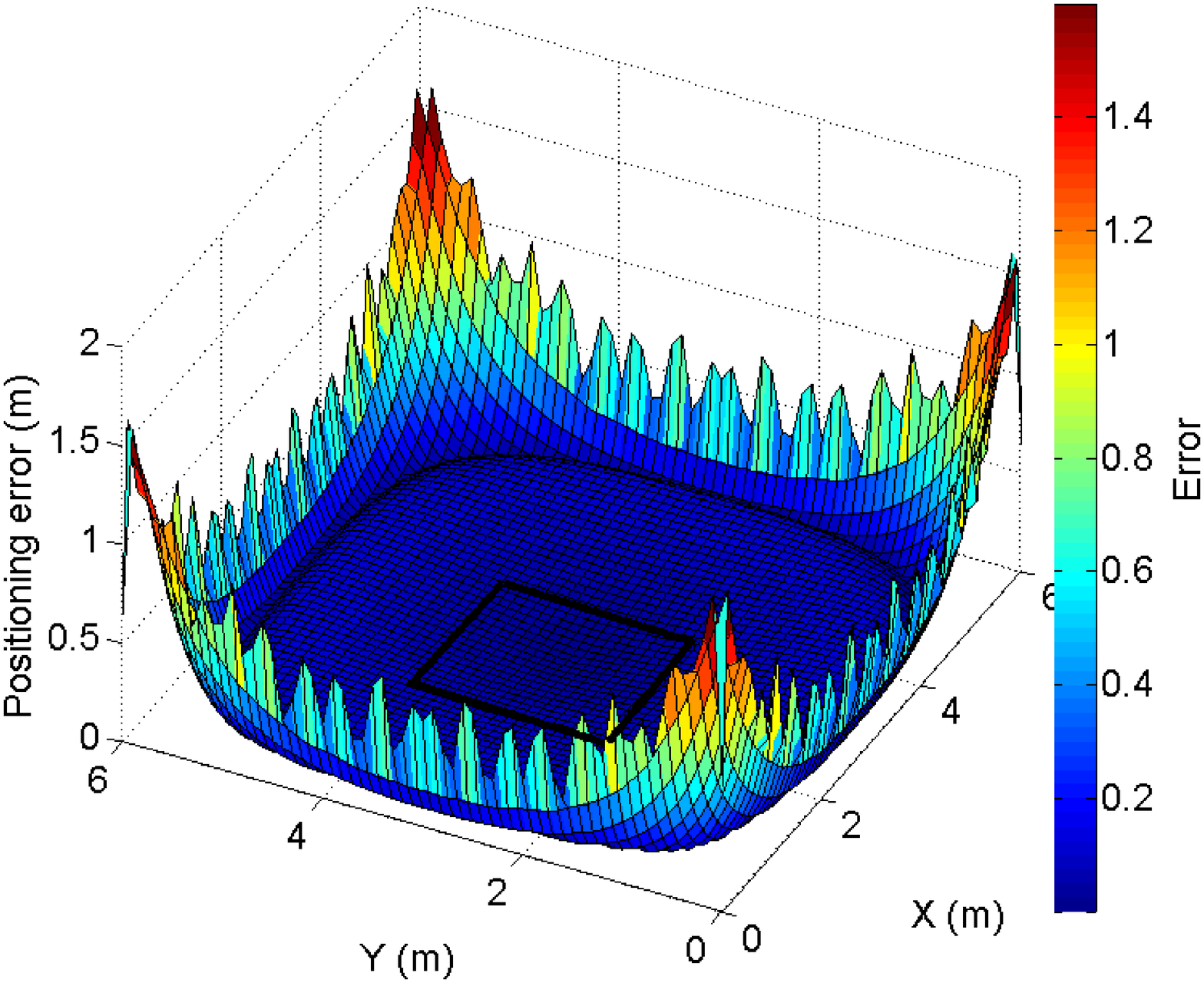}
\caption{Positioning error distribution for OFDM system with 64-QAM modulation, $N$ = 512 and $P_{t_e,k}$ = 5 dBm.}
\end{figure}
\begin{figure}
\centering
\includegraphics[width = 7cm, height =6cm]{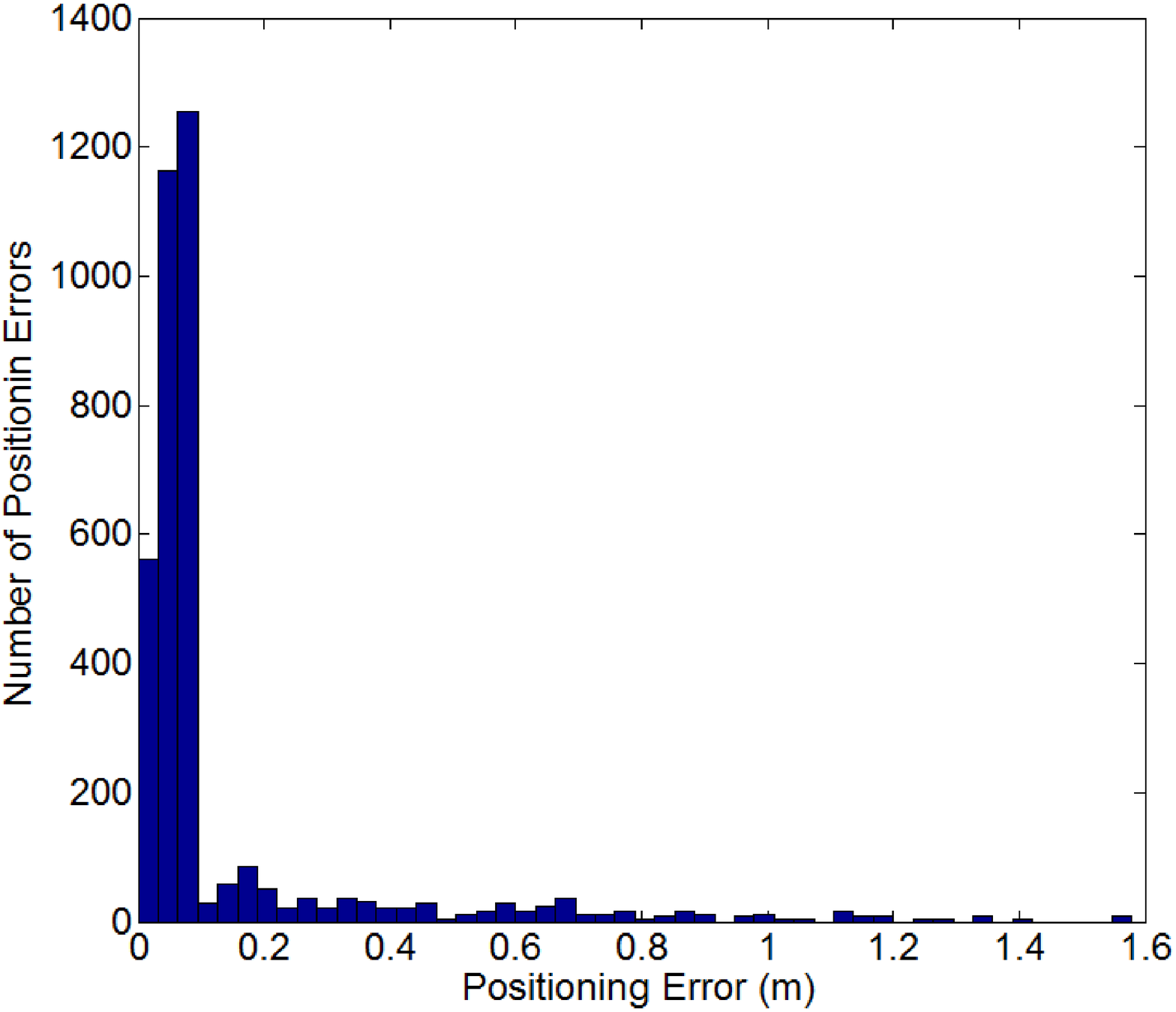}
\caption{Histogram of positioning errors for OFDM system with 16-QAM modulation, $N$ = 512 and $P_{t_e,k}$ = 5 dBm.}
\end{figure}
\begin{figure}
\centering
\includegraphics[width = 7cm, height = 6cm]{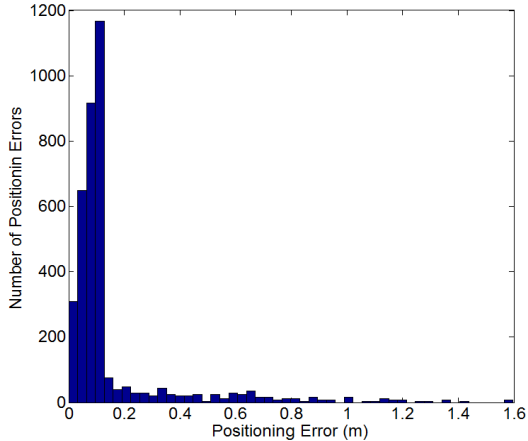}
\caption{Histogram of positioning errors for OFDM system with 64-QAM modulation, $N$ = 512 and $P_{t_e,k}$ = 5 dBm.}
\end{figure}
\subsection{Effect of number of subcarriers on the positioning accuracy}
Finally, we investigate the effect of number of total subcarriers (i.e. the FFT size) on the positioning accuracy. We consider an OFDM system with 4-QAM and SNR of 15 dB. Figs. 19-24 illustrate the positioning error distribution for different FFT sizes providing sufficiently narrow-banded sub-channels along with their corresponding error histograms. The total RMS errors are respectively calculated as 0.2905 m, 0.271 m and 0.2624 m for $N=64,256$ and 1024. Considering the results presented earlier for the FFT size of 512, it is observed that increasing the number of subcarriers results in a better positioning performance as it improves the estimation of the channel DC gain. However, the peak-to-average power ratio (PAPR) also increases with increasing the FFT size \cite{1}. Therefore, for sufficiently large values of $N$, the positioning performance is slightly degraded.

\begin{figure}
\centering
\includegraphics[width = 7cm, height = 6.5cm]{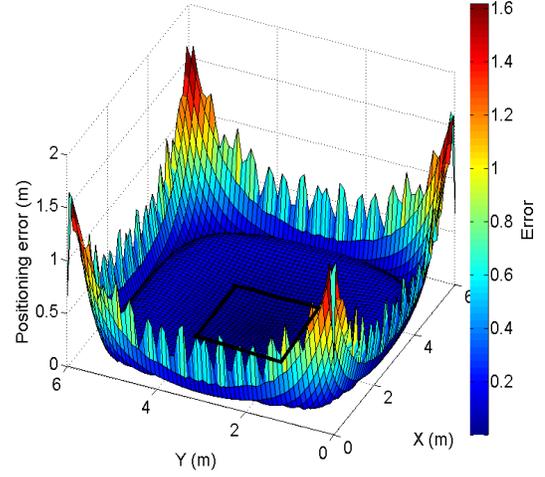}
\caption{Positioning error distribution for OFDM system with 4-QAM modulation, $N$ = 64 and $P_{t_e,k}$ = 5 dBm.}
\end{figure}
\begin{figure}
\centering
\includegraphics[width = 7cm, height = 6.5cm]{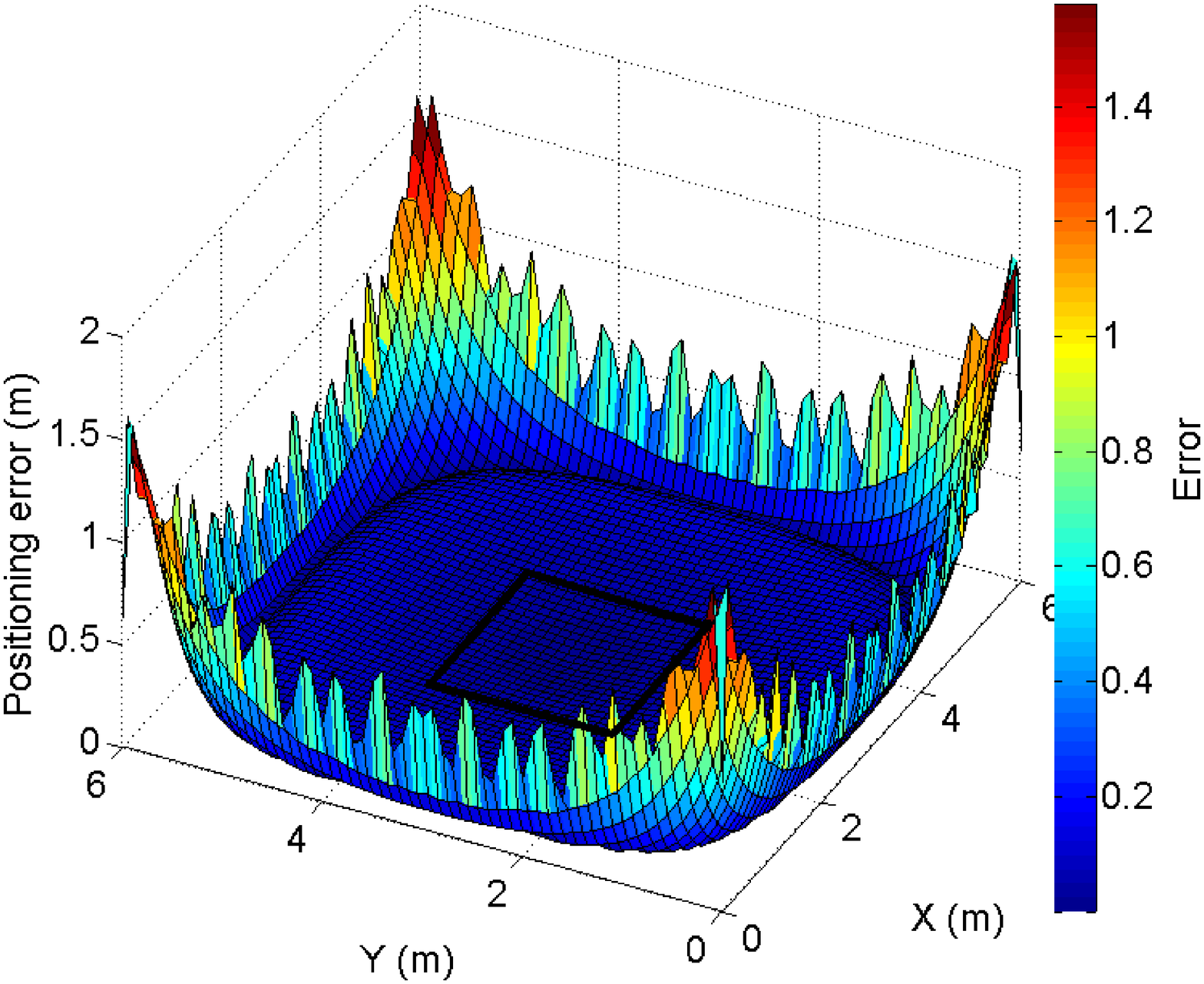}
\caption{Positioning error distribution for OFDM system with 4-QAM modulation, $N$ = 256 and $P_{t_e,k}$ = 5 dBm.}
\end{figure}
\begin{figure}
\centering
\includegraphics[width = 7cm, height = 6.5cm]{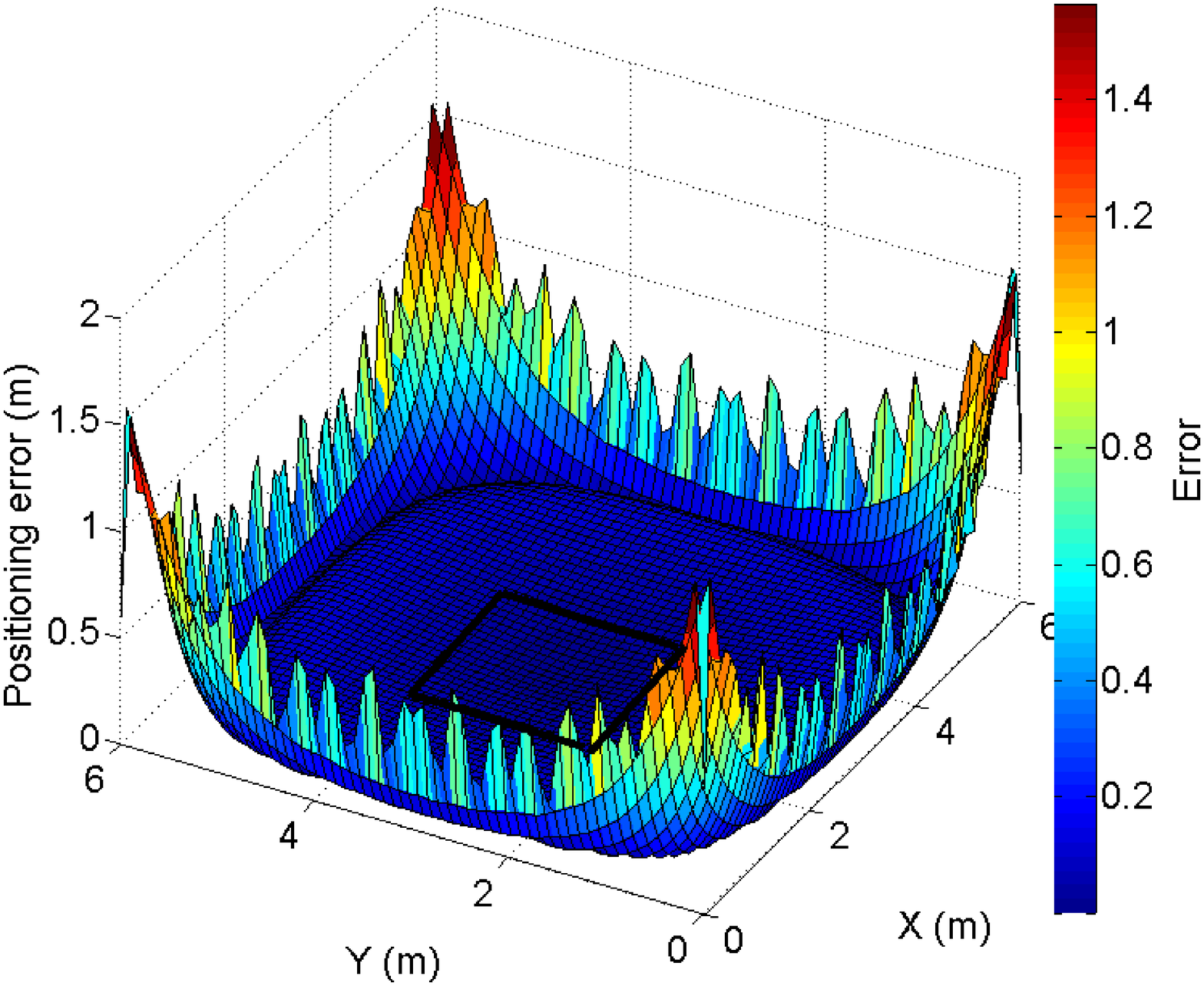}
\caption{Positioning error distribution for OFDM system with 4-QAM modulation, $N$ = 1024 and $P_{t_e,k}$ = 5 dBm.}
\end{figure}
\begin{figure}
\centering
\includegraphics[width = 7cm, height = 6.5cm]{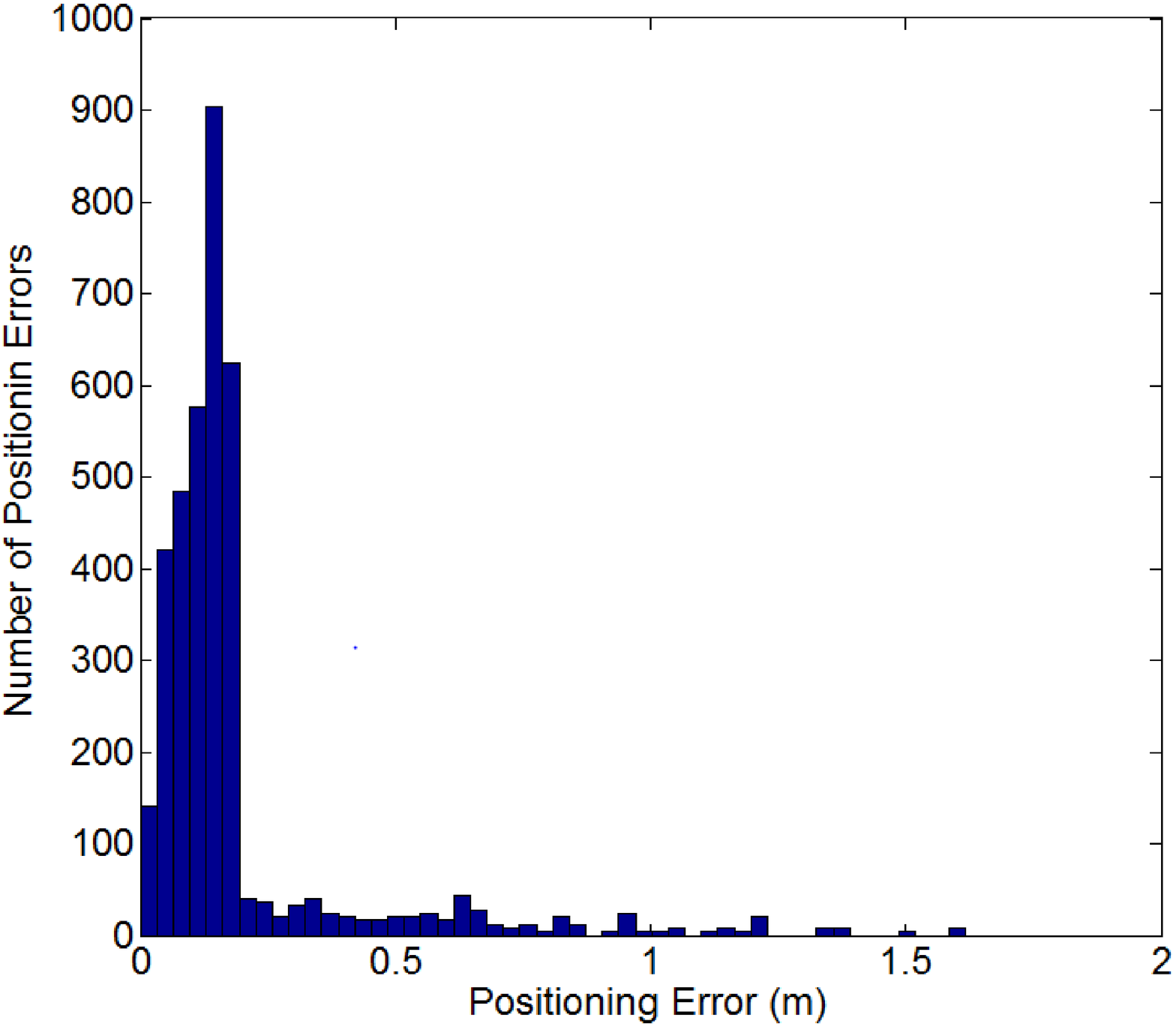}
\caption{Histogram of positioning errors for OFDM system with 4-QAM modulation, $N$ = 64 and $P_{t_e,k}$ = 5 dBm.}
\end{figure}
\begin{figure}
\centering
\includegraphics[width = 7cm, height = 6.5cm]{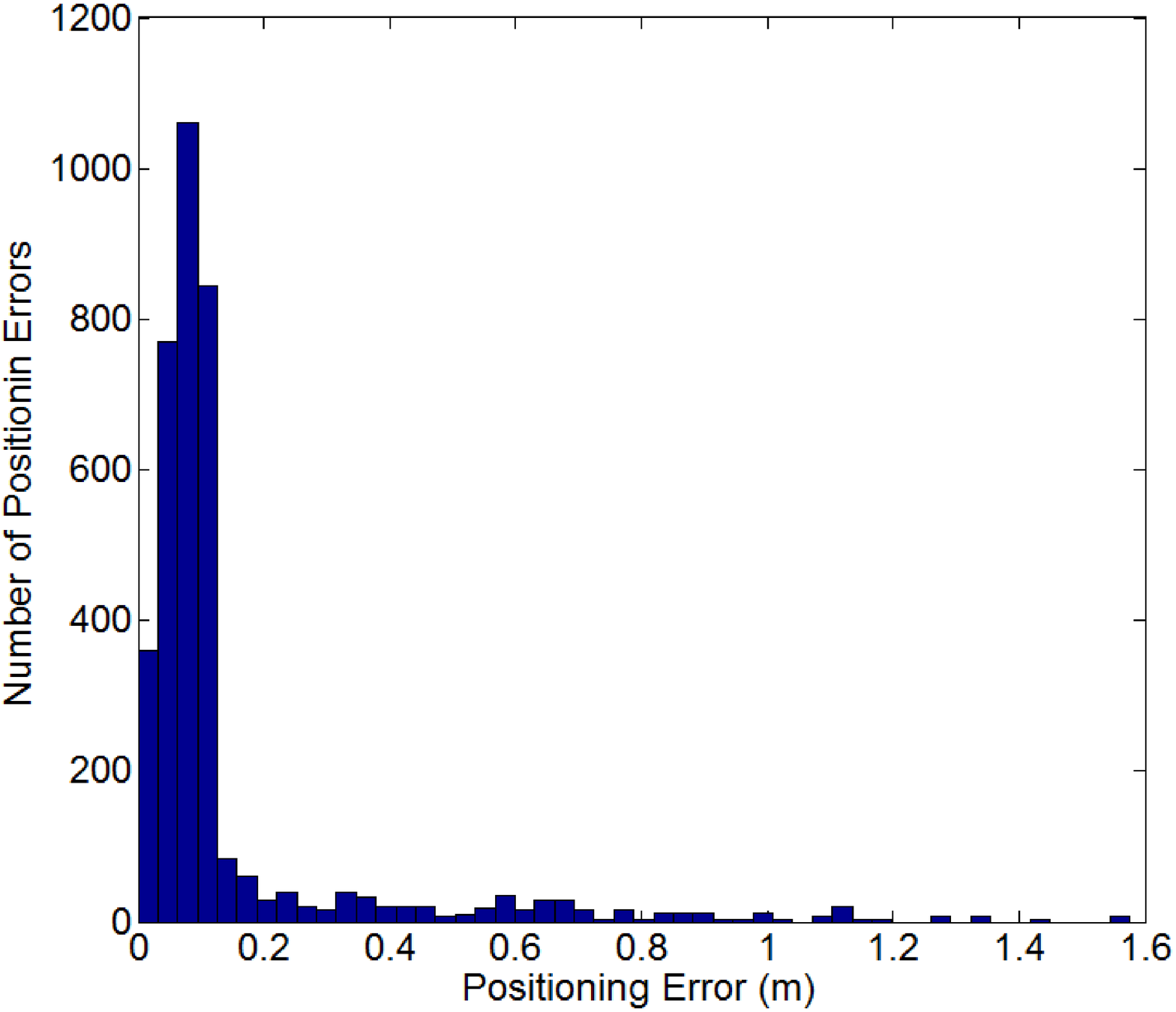}
\caption{Histogram of positioning errors for OFDM system with 4-QAM modulation, $N$ = 256 and $P_{t_e,k}$ = 5 dBm.}
\end{figure}
\begin{figure}
\centering
\includegraphics[width = 7cm, height = 6.5cm]{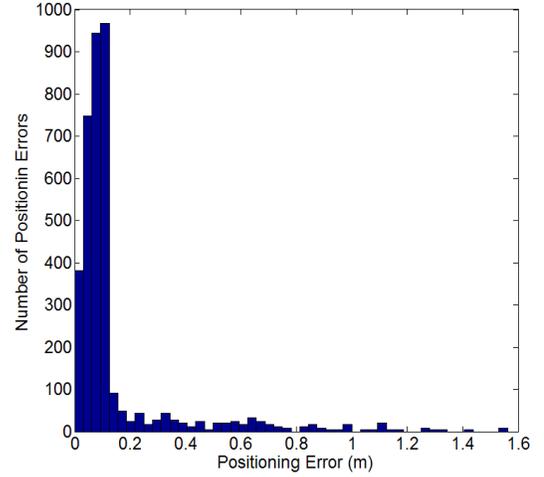}
\caption{Histogram of positioning errors for OFDM system with 4-QAM modulation, $N$ = 1024 and $P_{t_e,k}$ = 5 dBm.}
\end{figure}

\section{CONCLUSIONS}
In this paper, we have investigated and compared the positioning accuracy of IM/DD single- and multi-carrier modulation schemes for indoor VLC systems taking into account both nonlinear characteristics of LED and dispersive nature of optical wireless channel. Particularly, we have proposed an OFDM VLC system that can be used for both indoor positioning and communications. The training sequence used for synchronization has been adopted to estimate the channel DC gain. Lateration algorithm and the linear least squares estimation have been applied to calculate the receiver coordinates. We have shown the positioning error distribution over a typical room where the impulse response has been simulated employing CDMMC approach. Our results have demonstrated that the proposed OFDM system achieves an excellent accuracy and outperforms its OOK counterpart. Furthermore, the effect of different parameters on the positioning performance of the OFDM system have been investigated. We have shown that our proposed model unlike its conventional counterparts provides satisfactory performance at low SNR values and can be used for higher-order constellations as well.
\section*{Acknowledgement}
The authors would like to thank the National Science Foundation (NSF) ECCS directorate for their support of this work under Award \# 1201636, as well as Award \# 1160924, on the NSF “Center on Optical Wireless Applications (COWA– \href{http://cowa.psu.edu}{http://cowa.psu.edu})”
\balance
\bibliographystyle{IEEEtran}
\bibliography{ref}

\end{document}